\documentclass[11pt, onecolumn, draftclsnofoot]{IEEEtran}
\usepackage{amsmath,amssymb}
\usepackage[dvips]{graphicx}
\usepackage{amsfonts}
\usepackage[mathscr]{eucal}
\usepackage{latexsym}
\usepackage{amsthm}
\usepackage{exscale}
\usepackage[mathscr]{eucal}
\usepackage{bm}
\usepackage[dvipsnames]{color}
\usepackage{cases}
\usepackage{epsfig}
\usepackage[center,small]{caption}
\usepackage{algorithm}
\usepackage{algorithmic}
\usepackage[verbose,nospace,sort]{cite}
\usepackage{tabularx}
\usepackage{multirow}
\usepackage{balance}
\usepackage{url}
\scrollmode

\newtheorem{example}{Example}

\IEEEoverridecommandlockouts

\begin{document}
\title{Overcoming Endurance Issue: UAV-Enabled Communications with Proactive Caching}
\author{Xiaoli~Xu, Yong~Zeng, \emph{Member, IEEE,} Yong~Liang~Guan, \emph{Senior Member, IEEE}, and Rui~Zhang, \emph{Fellow, IEEE}\vspace{-8ex}
\thanks{X. Xu and Y. L. Guan are with the School of Electrical and Electronic Engineering, Nanyang Technological University, Singapore 639801 (email: \{xuxiaoli, eylguan\}@ntu.edu.sg).}
\thanks{Y. Zeng and R. Zhang are with the Department of Electrical and Computer Engineering, National University of Singapore, Singapore 117583 (email: \{elezeng, elezhang\}@nus.edu.sg).}
\thanks{This work was supported in part by the NTU-NXP Intelligent Transport System Test-Bed Living Lab Fund S15-1105-RF-LLF from the Economic Development Board, Singapore.
}}
\maketitle

\begin{abstract}
Wireless communication enabled by unmanned aerial vehicles (UAVs) has emerged as an appealing technology for many application scenarios in future wireless systems. However, the limited endurance of UAVs greatly hinders the practical implementation of UAV-enabled communications. To overcome this issue, this paper proposes a novel scheme for UAV-enabled communications by utilizing the promising technique of \emph{proactive caching} at the users. Specifically, we focus on content-centric communication systems, where a UAV is dispatched to serve a group of ground nodes (GNs) with random and asynchronous requests for files drawn from a given set.  With the proposed scheme, at the beginning of each operation period, the UAV pro-actively transmits the files to a subset of selected GNs that cooperatively cache all the files in the set. As a result, when requested, a file can be retrieved by each GN either directly from its local cache or from its nearest neighbor that has cached the file via device-to-device (D2D) communications. It is revealed that there exists a fundamental trade-off between the \emph{file caching cost}, which is the total time required for the UAV to transmit the files to their designated caching GNs, and the \emph{file retrieval cost}, which is the average time required for serving one file request. To characterize this trade-off, we formulate an optimization problem to minimize the weighted sum of the two costs, via jointly designing the file caching policy, the UAV trajectory and communication scheduling. As the formulated problem is NP-hard in general, we propose efficient algorithms to find high-quality approximate solutions for it.  Numerical results are provided to corroborate our study and show the great potential of proactive caching for overcoming the limited endurance issue  in UAV-enabled communications.
\end{abstract}
\vspace{-0.4cm}

\section{Introduction}
Traditionally, wireless communications are mainly designed with fixed terrestrial  infrastructure such as  ground base stations (BSs), access points, and relays. To meet the ever-increasing and highly diversified traffic demand cost-effectively, there have been fast growing interests in providing wireless connectivity from the sky, by using various aerial communication platforms such as balloons\footnote{Project Loon, Available online at: \url{https://x.company/loon/.}}, helikites \cite{Chandrasekharan2016}, and unmanned aerial vehicles (UAVs) \cite{Zeng2016} \cite{Yaliniz2016}. In particular, thanks to their fast deployment and controllable mobility, UAV-enabled/aided wireless communications have emerged as an appealing technology for many practical applications, such as terrestrial BS offloading \cite{lyuOffload}, emergency response and public safety \cite{Merwaday2015}, Internet-of-Things (IoT) communications \cite{MozaffariIoT}, \cite{ZhanIoT}, massive machine type communications \cite{Soorki2016}, etc.

Extensive research efforts have been recently devoted to UAV-enabled communications. In \cite{Zeng2016}, a general overview of UAV-enabled wireless communications is given, where three typical use cases are envisioned, namely UAV-aided ubiquitous coverage, UAV-aided relaying, and UAV-aided information dissemination and data collection. By employing UAVs as quasi-stationary aerial BSs, the UAV placement problem in two-dimensional (2D) or three-dimensional (3D) space has been extensively studied via exploiting the unique UAV-ground channel characteristics \cite{Hourani2014,BorYaliniz2016,Mozaffari2016,Lyu2017,Azari2016,Chen2017,He2017}. Moreover, another important line of work focuses on the UAV trajectory optimization \cite{UAVRelayZeng,ZengY2017,LyuZeng17,WuQQ2017,WuQQTPutDelay}, which fully exploits the additional design degrees of freedom introduced by the UAV mobility for communication performance enhancement.

 Despite all the promising benefits, UAV-enabled communications are also faced with new challenges. In particular, due to the practical size, weight, and power (SWAP) constraints, UAVs usually have very limited endurance or fly duration over the air. For example, most rotary-wing UAVs in the market typically have the maximum  endurance of about 30 minutes\footnote{DJI Phantom 4 Specs, Available online at: \url{https://www.drone-world.com/dji-phantom-4-specs/}}. This severely hinders the practical implementation of UAV-enabled communications. In particular, as the existing designs for UAV-enabled communications are mostly based on the conventional ``connection-centric" communication, where a communication link between the UAV and GN needs to be maintained for information transmission, a service interruption is caused when the UAV needs to be recalled for battery charging or swap. Some initial attempts have been made to prolong the UAV endurance or maximize the communication throughput given the limited energy, e.g., via energy-efficient trajectory designs \cite{ZengY2017}. However, the fundamental UAV endurance problem remains unresolved.

\begin{figure}[htb]
\centering
\includegraphics[scale=0.6]{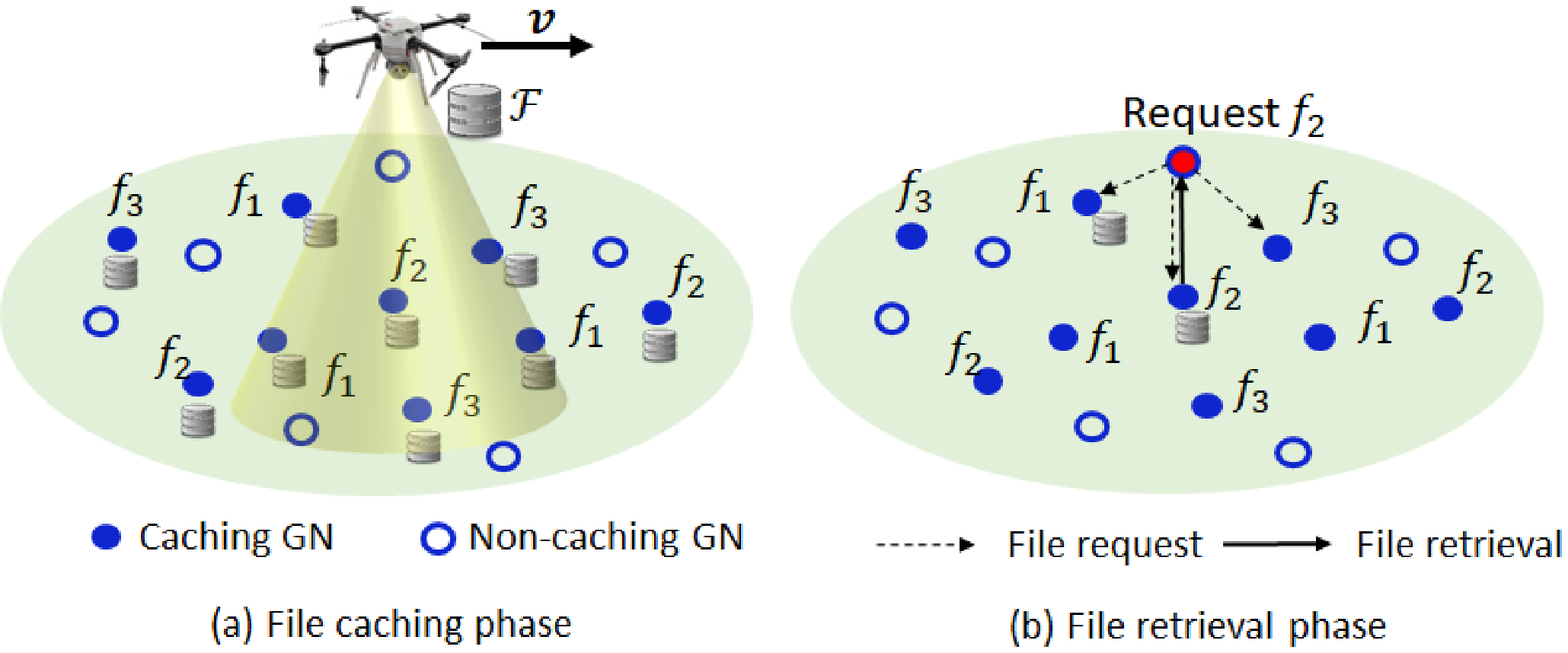}
\caption{Proposed scheme of UAV-enabled communication with proactive caching.}
\label{F:UAVCache}
\end{figure}

In this paper, we propose a novel scheme to overcome the UAV endurance issue,  by utilizing the promising technique of proactive caching at the GNs. Specifically, we focus on content-centric UAV-enabled wireless communication systems, where a UAV is dispatched to serve a group of GNs with random and {\it asynchronous} file requests, i.e., the same content may be requested by different GNs at different time. Note that ``content-centric" communications have many practical applications  nowadays, such as for video on-demand (VoD) streaming and  software download \cite{Cisco15}. As shown in Fig.~\ref{F:UAVCache}, the proposed scheme operates in a periodic manner, with each period consisting of two phases, namely the \emph{file caching phase} and the \emph{file retrieval phase}. In the file caching phase, the UAV pro-actively transmits each of the files from a given set of interest to a subset of selected GNs that cooperatively cache all the files. Next, in the file retrieval phase, each GN that has a file request can retrieve the file either directly from its own local cache or from its nearest neighbor that has cached the file via device-to-device (D2D) communications \cite{Guo2017}. For the proposed scheme, the UAV is only involved in the file caching phase. Thus, the required UAV operation time for each period only depends on how fast it can transmit the files to the selected caching GNs, instead of the random file request pattern of the GNs. This thus offers a promising solution to overcoming the issue of limited endurance for UAV-enabled communications. For instance, after completing the file caching, the UAV could return to the depot for  battery charging or swap, and yet without causing  any service interruption, thanks to the proactive file caching and D2D file sharing by  the GNs. It is worth noting  that the proposed scheme is fundamentally different from the UAV caching technique studied in \cite{UAVCacheSaad}, where the files are cached at the UAVs (instead of at GNs) based on the predicted content request distribution and mobility pattern of the users. In that work, the UAV needs to remain in the air for the entire period, thus  leaving the endurance issue unaddressed.

Note that caching has received significant interests recently in terrestrial cellular systems. By pre-loading the popular contents during off-peak hours into cellular BSs, mobile users, or even dedicated helper nodes \cite{Wang2014,Ji2016, Bastug14,Ali2014,Ji2015,Poularakis2016,TaoMX2016,ZhouTao2017,Golrezaei2013,Blaszczyszyn2015}, caching offers a promising approach to alleviate the backhaul congestion issue and reduce latency in cellular networks. In this paper, caching at the GNs is exploited  as a new means to overcoming the endurance issue for the emerging UAV-enabled communications, which, to the best of our knowledge, has not been reported in prior work.

Different from the caching in cellular networks, the file caching phase of the proposed scheme is via the UAV-to-GN wireless links, whose quality critically depends on their distances and thus the UAV trajectory over time. Therefore, the caching policy, which includes the decisions on who are the caching GNs and which files to be cached at each caching GN, should be jointly designed with the UAV trajectory and the UAV file transmission scheduling. Furthermore, there exists a fundamental trade-off between the {\it file caching cost}, which is defined as the total time required for the UAV to complete the transmission of all files to the selected caching GNs,  and the {\it file retrieval cost}, which is the average time required for serving one file request \cite{kangasharju2002}. {Note that the file caching cost is ignored in \cite{kangasharju2002}. However, due to the limited endurance of UAV, the file caching strategy should be carefully designed to minimize the caching cost by UAV as well. } Intuitively, with the given finite storage capacity at each caching GN, the file retrieval cost in general decreases if more files are cached to the GNs. However, this will incur higher file caching cost, since more time is required for the UAV to complete the file transmissions to the designated caching GNs. Furthermore, from the perspective of reducing the file retrieval cost, it is desirable to cache each file to those GNs that are well separated geographically, so that it is more likely for a GN  to find the requested file in a nearby caching GN if the file is not found in its local cache. However, this will increase the file caching cost since the UAV needs to travel longer distance in order to transmit the same file to those more separated caching GNs. {In this paper, we will investigate this new fundamental trade-off in detail, by jointly designing the caching strategy, UAV trajectory and file scheduling. Note that the problem we consider is fundamentally different from that studied in \cite{Zeng2017} where the UAV trajectory is optimized for broadcasting one common file to all the GNs. The file caching strategy and the file scheduling at the UAV, which are essential design parameters for our proposed scheme,  are irrelevant to the simple broadcasting problem considered in \cite{Zeng2017}.}

The main contributions of this paper are summarized as follows.
\begin{itemize}
\item{Firstly, we propose a novel scheme for UAV-enabled communication by utilizing proactive caching at the GNs to overcome the endurance issue. Furthermore, to characterize the fundamental trade-off between file caching and file retrieval costs, we formulate an optimization problem to minimize the weighted sum of the two costs via jointly optimizing the caching policy, UAV trajectory and communication scheduling.}
\item{Secondly, as the formulated optimization problem is NP-hard, we propose an efficient greedy based approach to find high-quality approximate solutions for it. The proposed algorithm starts from the case with no file cached, and then adds each caching file sequentially that leads to the maximum reduction in weighted sum cost. To minimize the file caching cost for any given caching policy, the UAV trajectory is designed by optimizing the waypoints based on the concept of virtual base station (VBS) placement, together with the linear programming (LP) for the speed optimization.}
\item{Thirdly, to further reduce the complexity of the proposed greedy solution, we propose an efficient approximation for estimating the file caching cost directly, without the need of solving  trajectory optimization problem at each iteration. With this low-complexity scheme, the UAV trajectory and communication  scheduling only needs to be optimized once  after determining the caching policy.}
\item{Lastly, extensive numerical results are provided to validate the performance of the proposed scheme and illustrate the trade-off between the file caching and retrieval costs. Furthermore, as compared to the benchmark schemes with separate caching policy and UAV trajectory designs, it is shown that the proposed scheme with joint caching and trajectory optimization achieves significant performance gains in terms of file caching and retrieval costs.}
\end{itemize}

The rest of this paper is organized as follows. Section~\ref{sec:model} presents the system model and proposes the novel scheme for UAV-enabled communication with caching. The problem formulation is also given in this section. In Section~\ref{sec:solution}, a greedy based solution is proposed for the formulated problem. Furthermore, a low-complexity scheme is proposed to further reduce the complexity of the proposed greedy solution by estimating the file caching cost directly. Section~\ref{sec:numerical} provides the numerical results, and finally we conclude this paper in Section~\ref{sec:con}.

\emph{Notations:} In this paper, scalars are denoted by italic letters. Boldface lower-case and upper-case letters denote vectors and matrices, respectively. $\mathbb{R}^{M\times 1}$ denotes the space of $M$-dimensional real-valued vectors. $\mathbb{Z}$ represents the set of non-negative integers. For a vector $\mathbf{a}$, $\lVert\mathbf{a}\rVert$ represents its Euclidean norm. $\log_2(\cdot)$ denotes the logarithm with base 2.  $\mathbb{E}[\cdot]$ denotes the statistical expectation and $\Pr(\cdot)$ represents the probability. For a time-dependent function $\mathbf{q}(t)$, $\dot{\mathbf q}(t)$ denotes the first-order derivative with respect to time $t$. For a set $\mathcal{K}$, $|\mathcal{K}|$ denotes its cardinality. For two sets $\mathcal{K}_1$ and $\mathcal{K}_2$, $\mathcal{K}_1\subset\mathcal{K}_2$ denotes that $\mathcal{K}_1$ is a subset of $\mathcal{K}_2$.  $\mathcal{K}_1\bigcup\mathcal{K}_2$, $\mathcal{K}_1\bigcap\mathcal{K}_2$ and $\mathcal{K}_1\setminus\mathcal{K}_2$ denote the union, intersection and set difference of $\mathcal{K}_1$ and $\mathcal{K}_2$, respectively.

\section{System Model and Problem Formulation}\label{sec:model}
 We consider a UAV-enabled wireless communication system, where a UAV is dispatched to serve a group of  $K$ GNs. The horizontal location of GN $k$ is denoted as $\mathbf{w}_k\in\mathbb{R}^{2\times 1},k\in\{1,2,...,K\}$. Different from the existing literature that mostly considers the traditional ``connection-centric" UAV-enabled communications \cite{Zeng2016,Yaliniz2016,lyuOffload,Merwaday2015,MozaffariIoT,ZhanIoT,Soorki2016,Hourani2014,BorYaliniz2016,Mozaffari2016,Lyu2017,Azari2016,Chen2017,He2017,UAVRelayZeng,ZengY2017,LyuZeng17,WuQQ2017,WuQQTPutDelay}, we consider the ``content-centric" UAV-enabled communications. Specifically, we assume that within each period of duration $T_{p}$ seconds, the $K$ GNs are interested in the same set of $N$ files, which are denoted as $\mathcal{F}=\{f_1,...,f_N\}$. 
  Note that in practice, $T_p$ depends on how fast the file library needs to be updated, which is usually at relatively large time scale (say one day). {We assume that the probability for GN $k$ to request file $f_n$ is denoted by $P_f^{(k)}(n)$, where $k=1,...K, n=1,...,N$ and $0\leq P_f^{(k)}(n)\leq 1$. For the special case when all the GNs have the same interest on the files, the file request probability can be represented by the file popularity, such as the Zipf distribution given by}
\begin{align}
P_f^{(k)}(n)=\frac{1/n^{\kappa}}{\sum_{n}1/n^{\kappa}},\  n=1,...,N, \forall k \label{eq:popularity}
\end{align}
where $\kappa$ represents the skewness of the distribution and usually takes values in $[0.5, 1.5]$ \cite{Breslau1999}. For the special case when $\kappa=0$, all the files are of equal popularity. As $\kappa$ increases, the popularity of different files becomes more diverged.


In practice, different GNs may request the same file at different time. A straightforward  way to satisfy such asynchronous file requests on demand  is via direct UAV-GN transmission. However, such a scheme requires the UAV to remain over the air for all time, similar to the conventional terrestrial BSs at fixed locations on the ground. However, in practice, UAVs have limited endurance and thus this scheme is  practically infeasible.  To overcome this issue,  we propose a novel scheme for UAV-enabled wireless communication based on proactive caching by the GNs.

\subsection{Benchmark Scheme: Direct UAV-GN File Transmission Without Caching}
\begin{figure}[htb]
\centering
\includegraphics[scale=0.6]{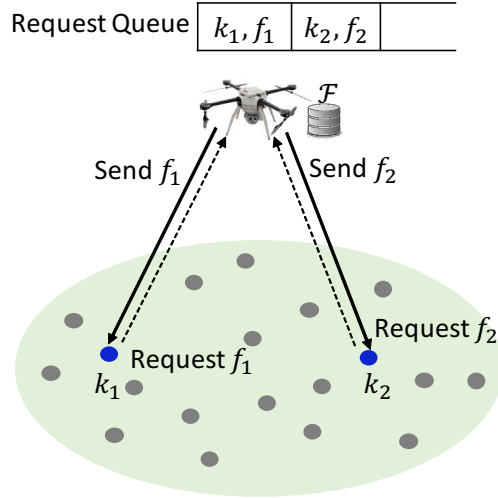}
\caption{Benchmark scheme using direct UAV-GN file transmission without caching.}
\label{F:benchmark}
\end{figure}

We first consider a benchmark scheme with direct UAV-GN file transmission without caching, where the UAV resembles a conventional ground BS,  but hovers above a certain location (e.g., the geometric center of the locations of all GNs) or flies along some optimized trajectory to directly serve the file request on demand.  As illustrated in Fig.~\ref{F:benchmark}, the file requests (including the requesting GN and the index of the requested file) from the GNs are put in a request queue at the UAV  based on their generated time. All such requests are sequentially served by the UAV via direct file transmissions. In practice, the GNs may have the file request at any time. Thus the UAV needs to remain in the air throughout the mission operation time (say several hours or even a day). This is challenging to be practically implemented  due to the limited UAV on-board energy and hence endurance. Moreover, such a scheme is also quite inefficient when the file requests from the GNs are highly sporadic, for which the UAV needs to remain in the air even when there is temporarily no file request. Therefore, in the following, we propose a new scheme for UAV-enabled communication based on the promising technique of proactive caching at the GNs.

\subsection{Proposed Scheme with Proactive Caching}
With the proposed scheme, each GN is assumed to have a cache with storage capacity of $Q$ files. As illustrated in Fig.~\ref{F:UAVCache}, the proposed scheme consists of two phases: {\it file caching phase} and {\it file retrieval phase}, explained as follows.

\begin{itemize}
\item{\emph{File caching phase}: The file caching phase occurs at the beginning of each operation period, for which the UAV selects a subset of the $K$ GNs to pro-actively cache the $N$ files. To ensure non-zero probability of successful file retrieval for all files, each of the $N$ files should  be cached by at least one of the selected caching GNs. Furthermore, due to the limited storage capacity at each GN and the incurred file caching cost by transmitting the files from the UAV to their designated caching GNs, the files to be cached at each caching GN should be carefully optimized, jointly with the UAV trajectory and the file transmission scheduling from the UAV to the caching GNs.  }
\item{\emph{File retrieval phase}: With all the $N$ files cooperatively cached by the selected caching GNs, the file retrieval phase then only involves the D2D communication among the GNs. In this case, each file request could be served by considering two possible scenarios.  If the requested file is already cached locally at the requesting GN itself, it can then be simply retrieved from the local cache; otherwise, the file requesting GN will broadcast the file request, and those caching GNs who have cached the requested file will respond. We assume that the GN will then download the file from the nearest GN that has cached this file via D2D communication.}
\end{itemize}

Note that with the proposed scheme, the UAV is only involved in the file caching phase. Thus, its operation time/cost at each period only depends on how fast the UAV can complete the transmission of all files to their respectively selected caching GNs, thus independent  of the random file request pattern of the GNs. Thus, regardless of the period duration $T_p$, the proposed scheme only requires that the UAV's endurance to be greater than the duration of the file caching phase, since the UAV can then replenish its energy during the file retrieval phase.


In the following, the file caching and retrieval phases are modeled in detail.

\subsubsection{File Caching}
We denote by $X$ in bits the size of each of the $N$ files\footnote{For simplicity, we assume that all files have the same size $X$, while the results of this paper can be extended to the case of unequal file sizes with only minor modification.}. Furthermore, we assume that each file is divided into $\bar{Y}=X/R_p$ packets,  with $R_p$ denoting the packet size in bits. We further assume that a packet-level erasure correction code such as the fountain code \cite{Fountain} or random linear code \cite{Ho06} is applied for each file, so that a file can be recovered from {\it any} $Y=(1+\epsilon)\bar{Y}$ coded packets, where $\epsilon\ll 1$ is the coding overhead.

For the file caching phase,  the UAV needs to determine the file caching policy, which includes the subset of the $K$ GNs for file caching as well as a subset of the $N$ files to be cached at each selected caching GN.  This can be mathematically represented by  the $KN$ binary indication variables $I_{k,n}$ as
\begin{align}
I_{k,n}=\begin{cases}
1, & \textnormal{file $f_n$ is cached at GN $k$} \\
0, & \textnormal{otherwise.}
\end{cases}
\end{align}
Due to the storage limitation, the total number of files cached at each GN should not exceed its storage capacity $Q$, i.e.,
\begin{align}
\sum_{n=1}^{N}I_{k,n}\leq Q,\  k=1,...,K.
\end{align}
Furthermore, since each file should be cached by at least one GN to ensure non-zero successful file retrieval probability, we  have
\begin{align}
\sum_{k=1}^{K}I_{k,n}\geq 1,\  n=1,...,N.
\end{align}

With the file caching policy $\{I_{k,n}\}$ determined, the UAV needs to transmit the files to their designated caching GNs following certain trajectory and file transmission scheduling. Assume that the UAV flies at a constant altitude, which is denoted as $H$ in meter. Furthermore, denote by $\mathbf{q}(t)\in\mathbb{R}^{2\times 1}, 0\leq t\leq T_U$, the UAV's flying trajectory projected on the ground, where $T_U\ll T_{p}$ is the total time required for the UAV to complete the file caching transmission. Let $V_{\max}$ denote the maximum UAV speed in meter per second (m/s). We then have the constraint $\lVert\dot{\mathbf{q}}(t)\rVert\leq V_{\max}$. The time-dependent distance between the UAV and the GNs can then be written as
\begin{align}
d_k(t)=\sqrt{H^2+\lVert\mathbf{q}(t)-\mathbf{w}_{k}\rVert^2}, \  0\leq t\leq T_U, k=1,...,K.
\end{align}

For ease of exposition, the time horizon $T_U$ is discretized into $M$ equal time slots, i.e., $T_U=M\delta_t$, with $\delta_t$ denoting the elemental slot length such that the distances between the UAV and the GNs are approximately constant within each slot. As a rule of thumb, we may choose $\delta_t$ such that $\delta_t V_{\max}\ll H$. Then the UAV trajectory $\mathbf{q}(t)$ can be approximated by the $M$-length sequence $\{\mathbf{q}[m]\}_{m=1}^{M}$, where $\mathbf{q}[m]$ denotes the UAV's horizontal location at the $m$th time slot. Denote by $\delta_{d}\triangleq \delta_tV_{\max}$ the maximum traveling distance of the UAV for one slot duration $\delta_t$. The UAV speed constraint can then be discretized as
\begin{align}
\|\mathbf{q}[m]-\mathbf{q}[m-1]\|\leq \delta_d,\   m=2,...,M. \label{eq:P1Speed}
\end{align}
The time-dependent distance between the UAV and the GNs can be written as
\begin{align}
d_{k}[m]=\sqrt{H^2+\lVert\mathbf{q}[m]-\mathbf{w}_k\rVert^2},\   1\leq m\leq M, k=1,2,...,K,
\end{align}
where $\lVert\mathbf{q}[m]-\mathbf{w}_k\rVert$ is the horizontal distance.

Preliminary channel measurement results show that UAV-to-ground channel typically consists of strong LoS links \cite{Matolak2015}. Therefore, for simplicity in this paper we assume that the channel between the UAV and each GN  is dominated by the LoS link. As a result, the channel power gain from the UAV to GN $k$ at slot $m$ can be modeled as
\begin{align}
\beta_{k}^U[m]=\beta_0^Ud_{k}^{-2}[m]=\frac{\beta_0^U}{H^2+\lVert\mathbf{q}[m]-\mathbf{w}_k\rVert^{2}},
\end{align}
where $\beta_0^U$ denotes the UAV-to-ground channel power gain at the reference distance of $d_0=1$ m.

Denote by $P_U$ the transmission power of the UAV. The received signal-to-noise ratio (SNR) by GN $k$ at the $m$th time slot is given by
\begin{align}
\gamma_{k}[m]=\frac{P_U\beta_{k}^U[m]}{\sigma^2}=\frac{\gamma_0^U}{H^2+\lVert\mathbf{q}[m]-\mathbf{w}_k\rVert^{2}},\label{eq:UAVSNR}
\end{align}
where $\sigma^2$ is the additive white Gaussian noise (AWGN) power and $\gamma_0^U\triangleq\frac{P_U\beta_0^U}{\sigma^2}$ is the SNR at the reference distance of $d_0=1$ m.

We assume that the UAV's transmission rate throughout the file caching phase is fixed, which is denoted as $R_U$ in bits per second (b/s). As a result, the time required to complete one packet transmission is $t_p^U\triangleq\frac{R_p}{R_U}$. With the slot duration fixed as $\delta_t$, the total number of packets that can be transmitted by the UAV within each time slot is $L=\frac{\delta_t}{t_p^U}=\frac{\delta_tR_U}{R_p}$. For convenience, we assume that $L\geq 1$ is an integer. For each time slot $m$, define $J_{m,n}$ as the number of packets that are transmitted by the UAV for file $f_n$. At each time slot $m$, the total number of packets transmitted by the UAV cannot be larger than $L$. Therefore, we should have
\begin{align}
\sum_{n=1}^{N}J_{m,n}\leq L, \  m=1,...,M. \label{eq:P1Time}
\end{align}

With the UAV's transmission rate fixed as $R_U$, a packet sent by the UAV at slot $m$ can be successfully received by GN $k$ if and only if $\gamma_k[m]\geq \gamma_{\mathrm{th}}^U$, where $\gamma_{\mathrm{th}}^U\triangleq (2^{R_U/B_U}-1)\Gamma$ with $B_U$ denoting the channel bandwidth and $\Gamma$ denoting the SNR gap between a practical modulation and coding scheme and the theoretical Gaussian signaling. By using \eqref{eq:UAVSNR}, this condition is equivalent to that the horizontal distance between UAV and GN $k$ should be no greater than a certain threshold, i.e.,
$\lVert \mathbf{q}[m]-\mathbf{w}_k\rVert \leq D_U$, where
\begin{align}
D_U=\sqrt{\frac{\gamma_0^{U}}{\gamma_{\mathrm{th}}^{U}}-H^2}. \label{eq:UAVCoverage}
\end{align}

It is not difficult to see that the UAV transmission rate $R_U$ should be chosen such that $\gamma_{\mathrm{th}}^U\leq\gamma_0^U/H^2$.  For each GN $k$, define $\mathcal M_k\subseteq\{1,...,M\}$ as the subset of all time slots such that the horizontal distance between the UAV and GN $k$ is no greater than $D_U$,  i.e.,
\begin{align}
\mathcal{M}_k\triangleq\{m: \lVert\mathbf{q}[m]-\mathbf{w}_k\rVert\leq D_U\}. \label{eq:connectionSet}
\end{align}

We refer to $\mathcal M_k$ as the set of contacting time slots for GN $k$ with the UAV. Hence, the total number of coded packets that can be successfully received by node $k$ for file $f_n$ is given by $\sum_{m\in\mathcal{M}_k}J_{m,n}$. If GN $k$ is selected to cache file $f_n$, i.e., $I_{k,n}=1$, it needs to receive a total of $Y$ coded packets to recover $f_n$. Thus, the UAV trajectory $\mathbf{q}[m]$ and the file transmission scheduling $J_{m,n}$ should satisfy the following constraint,
\begin{align}
\sum_{m\in\mathcal{M}_k}J_{m,n}\geq Y, \forall \{(k,n): I_{k,n}=1\}. \label{eq:P1Decoding}
\end{align}

In this paper, we define the file caching cost $C_U$ as the total time required for the UAV to complete the dedicated file transmissions to the selected caching GNs, i.e.,
\begin{align}
C_U=T_U=M\delta_t. \label{eq:cachingCost}
\end{align}
Specifically, for any given file caching policy $\{I_{k,n}\}$, $C_U$ is the required file caching time such that there exists a feasible UAV trajectory $\{\mathbf{q}[m]\}$ and file transmission scheduling $\{J_{m,n}\}$ that satisfy the constraints \eqref{eq:P1Speed}, \eqref{eq:P1Time}, and \eqref{eq:P1Decoding}.


\subsubsection{File Retrieval}
Next, we consider the file retrieval phase. After the file caching phase with the policy specified by $\{I_{k,n}\}$, for each file $f_n$, denote by $\mathcal{K}_n$ the set of GNs that have cached file $f_n$, i.e., $\mathcal{K}_n\triangleq\{k: I_{k,n}=1\}$. In the file retrieval phase, when a GN $k$ requests any file $f_n$, there are two possible scenarios. In the first scenario, file $f_n$ is already cached by GN $k$ itself, i.e., $k\in\mathcal{K}_n$, then the file  can be retrieved  directly from its own local cache. In this case, the cost for file retrieval is essentially zero. Otherwise, when $k\notin \mathcal{K}_n$, GN $k$ will retrieve file $f_n$ from its nearest peer that has cached $f_n$ via D2D communication, which requires additional time/delay due to D2D transmissions and thus incurs a non-zero cost. For any pair of $k$ and $n$ such that $k\notin\mathcal{K}_n$, let the file retrieval distance for GN $k$ to retrieve $f_n$ be denoted as $D_{k,n}$. We have
\begin{align}
D_{k,n}=\min\{d_{kj}: j\in\mathcal{K}_n\}, \label{eq:downloadDist}
\end{align}
where $d_{kj}\triangleq\lVert\mathbf{w}_k-\mathbf{w}_j\rVert$ is the distance between GN $k$ and $j$.

As a result, the average channel power gain for GN $k$ to retrieve file $f_n$ can be modeled as
\begin{align}
\beta_{k,n}^G=\beta_0^GD_{k,n}^{-\alpha},\label{eq:pathloss}
\end{align}
where $\beta_0^G$ denotes the  channel power gain at the reference of $d_0=1$ m, and $\alpha\geq2$ is the path loss exponent for the D2D channels between GNs.

Denote by $R_G$ in b/s the transmission rate of each GN for the D2D file sharing phase. Therefore, the time required to complete one packet transmission is $t_p^G\triangleq\frac{R_p}{R_G}$. Different from the UAV-to-GT channels that are typically dominated by LoS links, the terrestrial channels between different GNs are usually subject to additional fading with random variations. We assume quasi-static fading channels, where the instantaneous channel coefficients between the GNs remain unchanged for each packet duration $t_p^G$, and may vary across different packets. Therefore, the instantaneous channel gains for GN $k$ to retrieve the $i$th packet for $f_n$ can be modeled as
\begin{align}
h_{k,n}[i]=\sqrt{\beta_{k,n}^G}g_{k,n}[i],
\end{align}
where $\beta_{k,n}^G$ is the distance-dependent path loss component given by \eqref{eq:pathloss}, $g_{k,n}[i]$ is a random variable with $\mathbb{E}\left[|g_{k,n}[i]|^2\right]=1$ accounting for the fading component of the terrestrial channels, which are assumed to be independently and identically distributed (i.i.d.). Without loss of generality, denote by $F(x)$ the complementary cumulative distribution function (ccdf) for the fading channel power, i.e., $F(x)\triangleq\Pr\left(|g_{k,n}[i]|^2\geq x\right)$.

Denote by $P_G$ the D2D transmit power for the file sharing phase. Then, the instantaneous SNR for GN $k$ to download the $i$th packet for file $f_n$ is given by
\begin{align}
\gamma_{k,n}^G[i]=\frac{P_G|h_{k,n}[i]|^2}{\sigma^2}=\frac{P_G\beta_{k,n}^G}{\sigma^2}|g_{k,n}[i]|^2=\frac{P_G\beta_{0}^G}{\sigma^2 D_{k,n}^{\alpha}}|g_{k,n}[i]|^2=\frac{\gamma_0^G}{D_{k,n}^{\alpha}}|g_{k,n}[i]|^2,
\end{align}
where $\sigma^2$ is the noise power and $\gamma_0^G\triangleq\frac{P_G\beta_0^G}{\sigma^2}$ is the average SNR at the reference distance $d_0=1$ m.

With the GNs' transmission rate fixed as $R_G$, a packet sent by the nearest caching GN of file $f_n$ to the requesting GN $k$ can be successfully received if the instantaneous SNR is no smaller than the threshold $\gamma_{\mathrm{th}}^{G}\triangleq (2^{R_G/B_G}-1)\Gamma$, where $B_G$ denotes the channel bandwidth for the D2D communications and $\Gamma$ is the SNR gap. Hence, the probability for GN $k$ to successfully receive the $i$th packet of file $f_n$ is given by
\begin{align}
p^{\mathrm{succ}}_{k,n}&=\Pr\left(\gamma_{k,n}^G[i]\geq \gamma_{\mathrm{th}}^{G}\right)\nonumber\\
&=\Pr\left(|g_{k,n}[i]|^2 \geq \frac{\gamma_{\mathrm{th}}^G}{\gamma_0^{G}}D_{k,n}^{\alpha}\right)\nonumber\\
&=F\left(\frac{\gamma_{\mathrm{th}}^G}{\gamma_0^{G}}D_{k,n}^{\alpha}\right). \label{eq:GNoutage}
\end{align}
Note that $F(x)$ is a decreasing function and hence the successful packet reception probability $p^{\mathrm{succ}}_{k,n}$ decreases with the file retrieval distance $D_{k,n}$, as expected.

Recall that to recover each file, the GN should successfully receive at least $Y$ coded packets.  For each given pair of GN $k$ and file $f_n$, we define the file retrieval cost $c_{k,n}$ as the expected number of required packet transmissions so that on average $Y$  packets are received by the file requesting node $k$. If $I_{k,n}=1$ such that $f_n$ is already available in its own cache, no packet transmission is needed and hence we have $c_{k,n}=0$. Otherwise, GN $k$ will retrieve file $f_n$ from its nearest GN  with each transmitted packet having success probability $p^{\mathrm{succ}}_{k,n}$. Therefore, we set $c_{k,n}=\frac{Y}{p^{\mathrm{succ}}_{k,n}}$, which measures the average number of transmitted packets (or transmission time normalized to one packet duration $t_p^G$) to ensure the successful decoding of file $f_n$.  In summary, with the caching policy $I_{k,n}$, the file retrieval cost for each given pair of GN $k$ and file $f_n$ is
\begin{align}
c_{k,n}=\begin{cases}
0, & I_{k,n}=1 \\
\frac{Y}{p^{\mathrm{succ}}_{k,n}},  & \textnormal{otherwise}.
\end{cases}\label{eq:retrieveTx}
\end{align}
{ Note that if a file is not cached at any GN, the result retrieval cost of this file is infinity for any GN according to \eqref{eq:retrieveTx}, which hence implies the constraint that all the files should be cached. In practical implementation, such constraint can be easily removed by defining a finite cost value for not caching a file. }

We define the average file retrieval cost $C_{G}$ as the average time required for serving one file request, where the average is taken over all the $K$ GNs and all $N$ files based on their popularity, i.e,
\begin{align}
C_G=\left(\frac{1}{K}\sum_{n=1}^{N}\sum_{k=1}^{K}P_f^{(k)}(n)c_{k,n}\right)t_p^G, \label{eq:RetrievalCost}
\end{align}
where $P_f^{(k)}(n)$ is the probability of GN $k$ requesting file $f_n$ as defined in \eqref{eq:popularity}. { Note that $C_G$ can also be interpreted as the average delay for a typical GN to retrieve a file given constant transmission rate between the GNs. } As more file requests are served via D2D communication, more files will be available at the GNs and hence the file retrieval cost may gradually reduce as the time goes. The cost defined in \eqref{eq:RetrievalCost} corresponds to the worst-case file retrieval cost at the beginning of the file retrieval phase, by only considering the file availability at the GNs after proactive caching enabled by the  UAV. { If the GNs can adjust their transmission rates  based on the communication distance, the file retrieval cost can also be defined as the average transmission time required for retrieving a file at the maximal achievable communication rate, which is clearly a increasing function of the distance between the requesting GN and its nearest caching GN with the file. }

\subsection{Problem Formulation}
Based on the above discussions, it can be seen that both the file caching cost $C_U$ defined in \eqref{eq:cachingCost} and average file retrieval cost $C_G$ defined in \eqref{eq:RetrievalCost} are closely coupled with the file caching policy $\{I_{k,n}\}$. Intuitively, the file retrieval cost decreases if more files are cached, due to the higher probability of finding the requested file in the local cache as well as the decreased retrieval distance on average when D2D file transmission is needed, as can be inferred from \eqref{eq:downloadDist}, \eqref{eq:GNoutage}, \eqref{eq:retrieveTx} and \eqref{eq:RetrievalCost}. However, this is usually achieved with increased file caching cost, since more time is required for the UAV to complete the file transmissions to the designated caching GNs. Moreover, to reduce the file retrieval cost, each file should be cached to those GNs that are as widely separated as possible given the same number of caching GNs, so as to reduce the maximum file retrieval distance when D2D file transmission is needed.  However, this will also increase the file caching cost since the UAV needs to transmit the same file to more distant GNs by traveling longer distance. Therefore, there exists a fundamental trade-off between file caching and file retrieval costs for the proposed scheme.

To characterize this new trade-off, we define a weighted sum cost as
\begin{align}
C_{\theta}&=(1-\theta) C_U+\theta C_G, \nonumber\\
&=(1-\theta) M\delta_t+\theta\left(\frac{ t_p^G}{K}\sum_{n=1}^{N}P_f(n)\sum_{k=1}^{K}c_{k,n}\right)\label{eq:totalCost}
\end{align}
where $0\leq \theta\leq 1$ is the weighting factor to balance between the average file retrieval cost $C_G$ and the file caching cost $C_U$.

The complete trade-off between $C_U$ and $C_G$ can be obtained by minimizing $C_{\theta}$ for different $\theta$ values, via jointly optimizing the design variables including the file caching policy $\{I_{k,n}\}$, UAV operation time $M$, UAV trajectory $\{\mathbf{q}[m]\}$, and file transmission scheduling $\{J_{m,n}\}$. For any given $\theta$, the problem can be formulated as
\begin{subequations}
\begin{align}
\textbf{P1: \quad } &\min_{M,\{I_{k,n},\mathbf{q}[m],J_{m,n}\}}C_{\theta} \\
\textnormal{s.t.,\quad } &{ M,J_{m,n}\in\mathbb{Z}, \mathbf{q}[m]\in\mathbb{R}},\label{eq:integer}\\
& I_{k,n}\in\{0,1\},\ k=1,...,K; n=1,...,N,\label{eq:binCon}\\
&\sum_{n=1}^{N}I_{k,n}\leq Q,\  k=1,...,K, \label{eq:StorageCon}\\
&\sum_{k=1}^{K}I_{k,n}\geq 1,\  n=1,...,N, \label{eq:allStored}\\
&\|\mathbf{q}[m]-\mathbf{q}[m-1]\|\leq \delta_d,\  m=2,...,M, \label{eq:SpeedCon}\\
&\sum_{m\in\mathcal M_k}J_{m,n}\geq Y,\  \forall \{(k,n): I_{k,n}=1\}, \label{eq:DecodingCon}\\
&\sum_{n=1}^{N}J_{m,n}\leq L,\  m=1,...,M, \label{eq:noPacketsSlot}
\end{align}
\end{subequations}
where $\eqref{eq:StorageCon}$ corresponds to the caching capacity constraint at each GN; \eqref{eq:allStored} ensures that each file is cached  by at least one GN; \eqref{eq:SpeedCon} corresponds to the maximum speed constraint of the UAV; \eqref{eq:DecodingCon} ensures that all the caching GNs receive sufficient number of packets for their designated  caching files from the UAV; and \eqref{eq:noPacketsSlot} specifies the maximum number of packets that can be transmitted by the UAV during each time slot.

\section{Proposed Solution}\label{sec:solution}
Before solving the general problem \textbf{P1}, we first consider its two extreme cases with $\theta=1$ and $\theta=0$, respectively,  to gain important insights. Then the general problem \textbf{P1} with arbitrary $\theta$ values is solved with a greedy algorithm.
\subsection{Minimizing File Retrieval Cost with $\theta=1$}\label{sec:theta1}
When $\theta=1$, \textbf{P1} reduces to minimizing the file retrieval cost $C_{G}$, while ignoring the corresponding file caching cost. It follows from \eqref{eq:RetrievalCost} that for any given file caching policy $\{I_{k,n}\}$, $C_G$ is independent of the UAV trajectory $\{\mathbf{q}[m]\}$ and file transmission scheduling $\{J_{m,n}\}$. Furthermore, by ignoring the constant terms in the cost function,  \textbf{P1} reduces to
\begin{subequations}
\begin{align}
\textbf{P1(a): \quad } &\min_{\{I_{k,n}\}}\sum_{n=1}^{N}P_f(n)\sum_{k=1}^{K}c_{k,n} \\
\textnormal{s.t.,\quad } & \eqref{eq:binCon}, \eqref{eq:StorageCon}, \eqref{eq:allStored}
\end{align}
\end{subequations}
Note that the optimization variable $I_{k,n}$ affects the cost function of \textbf{P1(a)} implicitly via $c_{k,n}$ given in \eqref{eq:retrieveTx}. In the ideal case when each GN has sufficiently large storage capacity, i.e., $Q\geq N$, it is not difficult to see that the optimal solution to \textbf{P1(a)} is $I_{k,n}=1, \forall (k, n)$, i.e., each GN will cache all the $N$ files. In this case, for each GN, all the files can be directly retrieved from their local caches once requested. It then follows from \eqref{eq:retrieveTx} that the optimal value of \textbf{P1(a)} is $C_G=0$. For the general case with $Q<N$, problem \textbf{P1(a)} is difficult to be solved optimally. In \cite{kangasharju2002},  a related problem is studied to optimize the  file caching policy to minimize the average file retrieval distance $\bar{D}\triangleq\frac{1}{K}\sum_{n=1}^{N}P_f(n)\sum_{k=1}^{K}D_{k,n}$, which has been proven to be NP-hard. Since the file retrieval cost $c_{k,n}$ is a deterministic and increasing function of the file retrieval distance $D_{k,n}$, it can be shown that  \textbf{P1(a)} is also NP-hard by following the similar proof in \cite{kangasharju2002}.

Therefore, by following a similar \emph{globally-greedy} heuristic approach in \cite{kangasharju2002}, an efficient approximate solution can be found for problem \textbf{P1(a)} for $Q<N$.  Specifically, at each step, the incremental reduction of the file retrieval cost for placing file $f_n$ at GN $k$  is calculated for all the GN-file pairs. Then, the GN-file pair that leads to the largest cost reduction is selected as a new cache placement.  This process iterates until the storages of all the GNs are filled. Note that if a file is not cached by any GN, the retrieval cost is infinity. Therefore, with the \emph{globally-greedy} approach that maximizes the cost reduction at each iteration, the constraint \eqref{eq:allStored} is automatically satisfied, as long as the problem is feasible with $KQ\geq N$.

\subsection{Minimizing File Caching Cost with $\theta=0$}\label{sec:theta0}
When $\theta=0$, we have $C_{\theta}=C_{U}$ and \textbf{P1} reduces to minimizing the UAV file caching cost $C_{U}=M\delta_t$ while ensuring that each file is cached by at least one GN, and ignoring the file retrieval cost. By discarding the constant terms, \textbf{P1} reduces to
\begin{subequations}
\begin{align}
\textbf{P1(b): \quad } &\min_{M,\{I_{k,n},\mathbf{q}[m],J_{m,n}\}}M \\
\textnormal{s.t.,\quad } & \eqref{eq:integer}-\eqref{eq:noPacketsSlot}.
\end{align}
\end{subequations}
For the ideal case when each GN has sufficiently large storage, i.e., $Q\geq N$, one GN is sufficient to cache all the $N$ files. In this case, it is not difficult to see that the optimal solution to \textbf{P1(b)}  is to cache all the files in one single arbitrarily selected GN $k^*$, i.e., $I_{k^*,n}=1,\forall n=1,...,N$ and $I_{k,n}=0, \forall k\neq k^*$. As such, the UAV only needs to hover above the selected GN $k^*$, i.e., $\mathbf{q}[m]=\mathbf{w}_{k^*}, m=1,...,M$. In this case, the minimum caching time is determined by the transmission time for sending all the files to the single GN selected, i.e., $M=YN$.

For the general case with $Q<N$, problem \textbf{P1(b)} is difficult to be optimally solved.  A heuristic solution to problem \textbf{P1(b)} can be obtained from the proposed solution for solving \textbf{P1} with general $\theta$ values in the next subsection.

\subsection{General Case with Arbitrary $\theta$}\label{sec:general}
Due to the implicit relationship between the file caching and retrieval costs with respect to the design variables, including the caching policy $\{I_{k,n}\}$, the UAV trajectory $\{\mathbf{q}[m]\}$ and the file scheduling $\{J_{m,n}\}$,  problem \textbf{P1} for arbitrary $\theta$ values is even more challenging to solve compared to the special cases discussed in the preceding subsections. Therefore, finding the optimal solution to \textbf{P1}  is difficult in general. In this subsection, we propose a \emph{greedy} algorithm for solving \textbf{P1} approximately by jointly minimizing the file caching and retrieval costs with weighting factor $0\leq\theta\leq 1$. The main idea of the proposed greedy approach is as follows: Instead of optimizing over all possible caching policy $\{I_{k,n}\}$, we start from the case where no file is cached, i.e., $I_{k,n}=0, \forall k,n$, and select the best GN-file pair that leads to the maximum reduction of the weighted sum cost $C_\theta$ in \textbf{P1} at each iteration.  This process continues until the storages of all the GNs are filled or when $C_\theta$ cannot be further reduced.

For notational convenience, with $K$ GNs and $N$ files to be cached, we define a set $\mathcal I$ containing all the GN-file pairs, i.e., $\mathcal{I}\triangleq\{(k,n): k=1,...,K; n=1,...,N\}$. The cardinality of $\mathcal I$ is thus $|\mathcal I|=KN$.
Furthermore, for any particular file caching policy $\{I_{k,n}\}$, we denote by $\mathcal{I}_S\triangleq\{(k,n): I_{k,n}=1\}$  the subset of $\mathcal{I}$ containing all the selected GN-file pairs $(k,n)$ such that file $f_n$ is cached at GN $k$. At each iteration, a GN-file pair $(k,n)$ is a candidate pair that could be selected for caching in the next step only if $(k,n)\not\in \mathcal{I}_S$ and the storage of GN $k$ has not been filled yet, i.e., $\sum_{n=1}^{N}I_{k,n}<Q$. Therefore, for any file caching policy $\{I_{k,n}\}$, we further denote by $\mathcal{I}_C\triangleq\{(k,n): I_{k,n}=0  \textnormal{ and }  \sum_{j=1}^{N}I_{k,j}<Q\}$ the set of candidate GN-file pairs, considering the current caching status specified by $\mathcal{I}_S$. Clearly, we have  $\mathcal{I}_S\bigcap\mathcal{I}_C=\emptyset$ and $\mathcal{I}_S\bigcup\mathcal{I}_C\subseteq \mathcal{I}$ for any particular file caching policy $\{I_{k,n}\}$.

With the proposed greedy approach, we start with an empty set $\mathcal I_S=\emptyset$, i.e., no file is cached, and at each step, select the best element in the candidate set $\mathcal I_C$ to move to $\mathcal I_S$. For notational convenience, we denote by $C_U(\mathcal{I}_S)$ the file caching cost for the UAV to transmit all the files to the designated caching GNs as specified in $\mathcal{I}_S$, and denote by $C_G(\mathcal{I}_S)$ the average file retrieval cost based on the file placement corresponding to $\mathcal{I}_S$, as defined in \eqref{eq:RetrievalCost}. The detailed procedures of the proposed greedy algorithm are described as follows:
\begin{enumerate}
\item{We start from the case without file placement by initializing $\mathcal{I}_S=\emptyset$ and $\mathcal{I}_C=\mathcal I$. The corresponding file caching and retrieval cost are $C_U(\mathcal{I}_S)=0$ and $C_G(\mathcal{I}_S)=\infty$ (see \eqref{eq:retrieveTx} and \eqref{eq:RetrievalCost}), respectively, and hence the initial weighted sum cost for $\theta\neq 0$ is $C_\theta=\infty$. }
\item{At each step with the existing file placement $\mathcal{I}_S$, we find the best GN-file pair, denoted as $(k^*,n^*)$, in the candidate set $\mathcal{I}_C$ such that the corresponding cost reduction with this new file caching is maximized. Specifically, denote the new caching policy after moving $(\hat{k},\hat{n})$ from $\mathcal{I}_C$ to $\mathcal{I}_S$ by $\hat{\mathcal I}_S$, i.e., $\hat{\mathcal I}_S\triangleq\mathcal{I}_S\bigcup\{(\hat{k},\hat{n})\}$. Then, the corresponding cost reduction of problem \textbf{P1} is
    \begin{align}
    \Delta C_{\theta}(\hat{k},\hat{n})=\theta\left(C_G(\mathcal{I}_S)-C_G(\hat{\mathcal I}_S)\right)-(1-\theta)\left(C_U(\hat{\mathcal I}_S)-C_U(\mathcal{I}_S)\right).\label{eq:netCostReduction}
    \end{align}
    The first term on the right hand side (RHS) of \eqref{eq:netCostReduction} represents the reduction of the file retrieval cost with the additional caching $(\hat{k},\hat{n})$, while the second term corresponds to the associated increase of the file caching cost.  Therefore, at each greedy step, the following optimization problem is solved
    \begin{subequations}
    \begin{align}
    \textbf{P2: \quad } &\max_{(\hat k,\hat n)\in \mathcal{I}_C,M,\{\mathbf{q}[m],J_{m,n}\}} \Delta C_{\theta}(\hat k,\hat n) \\
    \textnormal{s.t.,\quad }    &\|\mathbf{q}[m]-\mathbf{q}[m-1]\|\leq \delta_d,\  m=2,...,M, \label{eq:SpeedCon2}\\
    &\sum_{m\in\mathcal M_k}J_{m,n}\geq Y,\  \forall (k,n)\in\hat{\mathcal I}_S \\
    &\sum_{n=1}^{N}J_{m,n}\leq L,\  m=1,...,M. \label{eq:noPacketsSlot2}
    \end{align}
    \end{subequations}
    Note that as compared to the original problem \textbf{P1}, the file caching placement in \textbf{P2} is optimized in a greedy manner, where only one additional cache $(\hat k,\hat n)$ in the candidate set $\mathcal I_C$ is to be added to the current caching policy $\mathcal I_S$. It is worth noting that the storage constraint \eqref{eq:StorageCon} is guaranteed due to the definition of $\mathcal I_C$, while the constraint \eqref{eq:allStored} may not be necessarily satisfied for the solution to \textbf{P2} in the intermediate steps. However, it is not difficult to see that after the greedy algorithm converges, the constraint \eqref{eq:allStored} is also guaranteed, since $C_{\theta}$ is infinity if any file is not cached at all.  }
\item{Denote the obtained solution for the GN-file to \textbf{P2} as $(k^*, n^*)$. The subsets $\mathcal{I}_S$ and $\mathcal{I}_C$ are updated as $\mathcal{I}_S=\mathcal{I}_S\bigcup\{(k^*,n^*)\}$ and $\mathcal{I}_C=\mathcal{I}_C\setminus\{(k^*,n^*)\}$, respectively.
     Furthermore, if the storage at GN $k^*$ is used up after caching $f_{n^*}$, no more files can be cached at GN $k^*$ in the following steps. Therefore, the set of candidate GN-file pair $\mathcal{I}_C$ for subsequent greedy steps is updated as
    \begin{align}
    \mathcal{I}_C=\begin{cases}\mathcal{I}_C\setminus\{(k^*,j),j=1,...,N\}, & \textnormal{if }\sum_{j=1}^{N}I_{k^*,j}=Q\\
    \mathcal{I}_C\setminus\{(k^*,n^*)\}, & \textnormal{otherwise.}\end{cases} \label{eq:updateRemain}
    \end{align}
    Note that in the first case of \eqref{eq:updateRemain}, all GN-file pairs containing GN $k^*$ are excluded from $\mathcal I_C$, whereas in the second case, only the pair $(k^*, n^*)$ is excluded.}
\item{Repeat step (2)-(3) until there is no remaining feasible GN-file pair, i.e., $\mathcal{I}_C=\emptyset$, or the incremental cost reduction becomes negligible. }
\end{enumerate}

The remaining task for the proposed greedy scheme is then to solve \textbf{P2}. For any chosen candidate GN-file pair $(\hat k,\hat n)\in\mathcal{I}_C$, \textbf{P2} reduces to a joint UAV trajectory and transmission scheduling optimization problem formulated as
\begin{subequations}
\begin{align}
\textbf{P3: \quad } &\max_{M,\{\mathbf{q}[m],J_{m,n}\}} \Delta C_{\theta}(\hat k,\hat n) \\
\textnormal{s.t.,\quad } & \eqref{eq:SpeedCon2}-\eqref{eq:noPacketsSlot2}
\end{align}
\end{subequations}
Therefore, \textbf{P2} can be solved by simply comparing all the $|\mathcal I_C|$ possible values to \textbf{P3}, each corresponding to one feasible $(\hat k,\hat n)$ in $\mathcal I_C$. Note that with $\mathcal I_S$ and $\hat{\mathcal I}_S$ given, the reduction of the file retrieval cost $\theta (C_G(\mathcal I_S)-C_G(\hat{\mathcal{I}}_S))$ is a constant value that can be directly calculated from \eqref{eq:RetrievalCost}. Hence, solving \textbf{P3} is equivalent to minimizing the file caching cost $C_U(\hat{\mathcal{I}}_S)$ via optimizing the UAV trajectory and file transmission scheduling given the caching placement $\hat{\mathcal I}_S$, which is formulated as
\begin{subequations}
\begin{align}
\textbf{P3(a): \quad } &\min_{M,\{\mathbf{q}[m],J_{m,n}\}}C_{U}(\hat{\mathcal I}_S) \\
\textnormal{s.t.,\quad }
&\|\mathbf{q}[m]-\mathbf{q}[m-1]\|\leq \delta_d,\  m=2,...,M, \label{eq:SpeedCon3}\\
&\sum_{m\in\mathcal M_k}J_{m,n}\geq Y,\  \forall (k,n)\in \hat{\mathcal I}_S, \label{eq:DecodingCon3}\\
&\sum_{n=1}^{N}J_{m,n}\leq L,\  m=1,...,M. \label{eq:noPacketsSlot3}
\end{align}
\end{subequations}

It is worth noting that \textbf{P3(a)} is still a challenging problem since the set of contacting time slots $\mathcal{M}_k$ for each GN $k$ are intricately related with the UAV trajectory $\mathbf{q}[m]$, as given in \eqref{eq:connectionSet}.  For the special case when each file is only cached by one GN, \textbf{P3(a)} reduces to a problem for UAV-enabled unicast communication. On the other hand, when all the files are placed at all the GNs, i.e., $I_{k,n}=1, \forall (k,n)\in \hat{\mathcal I}_S$, the problem reduces to the UAV-enabled multicast problem, which has been studied in \cite{Zeng2017}. Note that both the UAV unicast and multicast problems are NP-hard in general. Hence, it is generally challenging to find the optimal UAV trajectory and transmission scheduling solution to problem \textbf{P3(a)}. By following the similar approach as in \cite{Zeng2017},  we propose an efficient and approximate solution to \textbf{P3(a)} by firstly designing the UAV flying path via applying the concept of virtual base station (VBS) placement and convex optimization, and then finding the optimal UAV speed and file transmission scheduling with  the given path by solving a linear programming (LP) problem.

\subsubsection{Design of UAV Flying Path}
First, to satisfy the constraint \eqref{eq:DecodingCon3} of \textbf{P3(a)}, the UAV trajectory needs to be designed such that all the caching GNs, denoted as $\hat{\mathcal{K}}\triangleq\{k: \exists (k,n)\in\hat{\mathcal I}_S\}$, can effectively  communicate with the UAV, i.e., there exists at least one location along the UAV path for each caching GN such that its distance with UAV is no greater than the required threshold $D_U$. Given the UAV coverage radius $D_U$, this resembles the classical traveling salesman problem with neighborhood (TSPN). In the following, we provide an efficient path design for solving the TSPN problem based on the VBS placement and convex optimization techniques.

\begin{figure}[htb]
\centering
\includegraphics[scale=0.6]{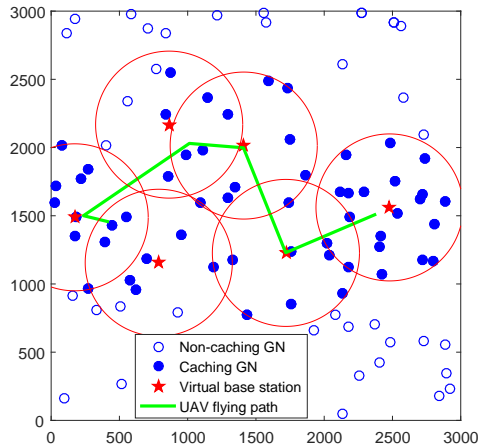}
\caption{An illustration of UAV path design with VBS placement and convex optimization.}
\label{F:VBS}
\end{figure}

Specifically, given the  locations of the caching GNs $\{\mathbf{w}_k, k\in\hat{\mathcal{K}}\}$ and the UAV coverage range $D_U$, we first solve the VBS placement problem, which aims to find a minimum number of VBSs and their respective locations, so that each caching GN is covered by at least one VBS. Several efficient algorithms in the literature can be applied for solving the VBS placement problem, where in this paper we adopt the spiral placement algorithm proposed in \cite{Lyu2017}. Let $G\leq |\hat{\mathcal K}|$ be the minimum number of VBSs obtained by applying the spiral  placement algorithm, and their locations are denoted as $\mathbf{v}_g\in\mathbb{R}^{2\times 1}, g=1,...,G$. An illustrative example for the VBS placement solution is shown in Fig.~\ref{F:VBS}.

With the locations of the VBSs found, the standard TSP algorithm can be applied to find the visiting order of the $G$ VBSs, which gives one feasible UAV path. However, the path can be further improved since it may not be necessary for the UAV to visit exactly the VBS locations \cite{Zeng2017}. To this end, convex optimization technique is further applied to optimize the set of way points. Specifically, with the VBSs and their visiting order determined,  the caching GNs in $\hat{\mathcal K}$ are essentially partitioned into $G$ ordered clusters $\hat{\mathcal{K}}_g, g=1,...,G$, where $\hat{\mathcal{K}}_g\subset \hat{\mathcal{K}}$ denotes the subset of caching GNs that must be covered by the $g$th VBS. Then, the starting and ending points of the UAV flying path intersecting with the $g$th cluster can be further optimized by solving a convex optimization problem, as described in \cite{Zeng2017}. By connecting the starting and ending points sequentially according to the order of the clusters visited, we obtain the complete UAV flying path, so that each caching GN is enabled to communicate with the UAV. An example of the obtained UAV flying path is shown in  Fig.~\ref{F:VBS}.

\subsubsection{UAV Speed and File Scheduling Optimization}\label{sec:scheduling}
With the UAV flying path determined, the locations along the path where each GN is covered by the UAV can be obtained based on \eqref{eq:connectionSet}. As a result, finding the UAV trajectory reduces to finding its instantaneous speed along the determined path, for which \textbf{P3(a)} can be cast as an  LP problem. Specifically, we discretize the obtained UAV flying path into $S$ segments, such that within each segment, the set of caching GNs that are in contact with the UAV remain unchanged. The length of the $s$th segment is denoted by $l_s, s=1,2,...,S$. The total UAV traveling distance is thus  $\sum_{i=1}^{S}l_i$. We further denote by $\rho_s$ the  time for the UAV to fly through the $s$th segment. The total file caching cost or the total UAV caching time is thus $C_U=\sum_{s=1}^{S}\rho_s$.  The speed constraint in \eqref{eq:SpeedCon3} is then equivalent to
\begin{align}
\rho_sV_{\max}\geq l_s, s=1,...,S.
\end{align}
Since the duration for the UAV sending one packet is $t_p^U$, the total number of packets that can be sent when it flies through the $s$th segment is $L_s=\rho_s/t_p^U$.  Then, the file transmission scheduling parameters $\{J_{m,n}\}$ can be equivalently transformed to $\{\tilde{J}_{s,n}\}$, where $\tilde{J}_{s,n}$ refers to the number of coded packets sent by the UAV for file $f_n$ during the $s$th path segment. Corresponding to \eqref{eq:noPacketsSlot3}, we have the following constraints on the total number of transmitted packets per segment,
\begin{align}
\sum_{n=1}^{N}\tilde{J}_{s,n}\leq L_s, s=1,...,S.
\end{align}
For each GN $k$, define $\mathcal{S}_k\subseteq\{1,...,S\}$ as the subset of the segments such that the GN $k$ can successfully receive the packet sent from the UAV, i.e., their horizontal distance being no greater than $D_U$. Then, the file recovery constraint in \eqref{eq:DecodingCon3} can be recast as
\begin{align}
\sum_{s\in\mathcal{S}_k}\tilde{J}_{s,n}\geq Y, \forall \{(k,n): I_{k,n}=1\}.
\end{align}

Therefore, with any given UAV flying path, problem \textbf{P3(a)} reduces to the following LP problem, which can be efficiently solved by standard convex optimization techniques.
\begin{subequations}
\begin{align}
\textbf{P4: \quad } &\min_{\{\rho_i,\tilde{J}_{s,n}\}}\sum_{s=1}^{S}\rho_s \\
\textnormal{s.t.,\quad }
&\rho_sV_{\max}\geq l_s,\  s=1,...,S, \label{eq:SpeedCon4}\\
&\sum_{s\in\mathcal{S}_k}\tilde{J}_{s,n}\geq Y,\  \forall \{(k,n): I_{k,n}=1\}, \label{eq:DecodingCon4}\\
&\sum_{n=1}^{N}\tilde{J}_{s,n}\leq L_s,\  s=1,...,S, \label{eq:noPacketsSlot4}
\end{align}
\end{subequations}
{
The pseudo-code for solving problem \textbf{P3(a)} is summarized in Algorithm~\ref{algP3a}. }
\begin{algorithm}

\caption{Algorithm for solving Problem \textbf{P3(a)}}
\label{algP3a}
\begin{algorithmic}[1]
\STATE{\textbf{Input: }$ \{\mathbf{w}_k,k\in\hat{\mathcal K}\},D_U,V_{\max}.$ }
\STATE{Find the location of the VBSs $\mathbf{v}_g, g=1,...,G$ to cover all caching GNs in $\hat{\mathcal K}$}
\STATE{Find the optimal visiting order of the $G$ VBSs using standard TSP algorithm.}
\STATE{Optimize the set of way points by solving convex optimization problem in \cite{Zeng2017}. }
\STATE{Divide the obtained flying path into segments, within which the UAV contact set remains unchanged.}
\STATE{Obtain the file scheduling and average flying speed within each segment by solving  \textbf{P4}.}
\STATE{\textbf{Output}: $ \{\mathbf{q}[m]\}$ and $\{J_{m,n}\}$}
\end{algorithmic}
\end{algorithm}

With problem \textbf{P3(a)} solved, the pseudo-code of the overall greedy algorithm proposed for problem \textbf{P1} is summarized in Algorithm~\ref{alg1}. Since one GN-file is selected at each iteration and each GN can cache at most $Q$ files, the total number of iterations of the proposed greedy algorithm is upper-bounded by $\min\{KN,KQ\}$, i.e., when all the files are placed at each GN or when all the GN storages are used up. Note that the actual number of iterations may be much smaller than this upper bound since the proposed approach may terminate when the incremental cost reduction is insignificant, i.e., $\Delta C_\theta <\epsilon$, where $\epsilon>0$ is a small positive parameter that controls the algorithm termination.
\begin{algorithm}
\caption{Greedy Algorithm for Problem \textbf{P1}}
\label{alg1}
\begin{algorithmic}[1]
\STATE{\textbf{Input: }$ \{\mathbf{w}_k,k=1,...,K\},\{P_f(n),n=1,...,N\},Q,D_U.$ }
\STATE{\textbf{Initialize:} Selected GN-file pairs $\mathcal{I}_S=\emptyset$, candidate GN-file pairs $\mathcal{I}_C=\mathcal{I}$, file retrieval cost $C_{G}(\mathcal{I}_S)=\infty$, file caching cost $C_{U}(\mathcal{I}_S)=0$ and $\Delta C_{\theta}=\infty$.}
\WHILE{$\Delta C_{\theta}>\epsilon$ and $\mathcal{I}_C\neq\emptyset$.}
\FOR{$(k,n)\in\mathcal{I}_C$}
\STATE{$\hat{\mathcal I}_S=\mathcal{I}_S\bigcup\{(k,n)\}.$}
\STATE{Calculate the file retrieval cost $C_G(\hat{\mathcal I}_S)$ from \eqref{eq:RetrievalCost}.}
\STATE{Solve \textbf{P3(a)} for the file caching cost $C_U(\hat{\mathcal I}_S)$.}
\STATE{Calculate the net cost reduction with \eqref{eq:netCostReduction}.}
\ENDFOR
\STATE{Denote the GN-file pair that leads to the highest cost reduction $\Delta C_{\theta}$ as $(k^*,n^*)$.}
\STATE{Update file retrieval cost $C_{G}(\mathcal{I}_S)=C_G(\mathcal{I}_S\bigcup\{(k^*,n^*)\})$.}
\STATE{Update file caching cost $C_{U}(\mathcal{I}_S)=C_U(\mathcal{I}_S\bigcup\{(k^*,n^*)\})$.}
\STATE{Update the selected GN-file pairs $\mathcal{I}=\mathcal{I}\bigcup\{(k^*,n^*)\}$.}
\STATE{Update the candidate set $\mathcal{I}_C$ as in \eqref{eq:updateRemain}.}
\ENDWHILE
\STATE{Solve \textbf{P3(a)} with $\mathcal{I}_S$ to obtain the UAV trajectory $\{\mathbf{q}[m]\}$ and file scheduling $\{J_{m,n}\}$.}
\STATE{\textbf{Output}: $\mathcal{I}_S, \{\mathbf{q}[m]\}$ and $\{J_{m,n}\}$}
\end{algorithmic}
\end{algorithm}

\subsection{Complexity Reduction with File Caching Cost Estimation}\label{sec:lowComp}
Note that at each iteration of Algorithm~\ref{alg1}, problem \textbf{P3(a)} needs to be solved for $|\mathcal{I}_C|$ times. In this subsection, the complexity of Algorithm~\ref{alg1} is further reduced via efficient estimation for the file caching cost directly, without the need for solving the optimization problem \textbf{P3(a)} as an  intermediate step. Specifically, for caching file $f_n$ at GN $k$, the file caching cost is estimated by assuming that the UAV will fly to the location above  GN $k$, and hover at  it until file $f_n$ is completely transmitted. Then, the estimated total file caching cost is the sum of the UAV flying time and the transmission time. Due to the broadcast nature of wireless communications,  all the neighbors of GN $k$ with distance no greater than the UAV coverage, denoted as $\mathcal{N}_k\triangleq\{k', d_{kk'}\leq D_U\}$, can overhear the file $f_n$. Hence, if we also need to cache file $f_n$ at any neighbor of GN $k$ in the subsequent iterations, no additional caching cost will be incurred due to the overhearing.

For notational convenience, we denote the set of GNs that have been visited by the UAV as $\mathcal{K}_v\subset\{1,...,K\}$. We further denote the overhearing status of all the GNs and all the files by the $KN$ binary indication function $O_{k,n}$ as
\begin{align}
O_{k,n}=
\begin{cases}
1, &\textnormal{GN $k$ has overheard file $f_n$}\\
0, &\textnormal{otherwise.}
\end{cases}
\end{align}

At the beginning of the proposed greedy approach, we have $\mathcal{K}_v=\emptyset$ and $O_{k,n}=0, \forall k,n$. At each iteration, we need to calculate the increase of file caching cost if we move the GN-file pair $(k,n)$ from the candidate GN-file pair set  $\mathcal{I}_C$ to the selected GN-file pair set $\mathcal{I}_S$, denoted as $\Delta C_U(k,n)\triangleq C_U(\hat{\mathcal{I}}_S)-C_U(\mathcal{I}_S)$, as part of the total cost reduction in \eqref{eq:netCostReduction}, where $\hat{\mathcal{I}}_S\triangleq{\mathcal{I}}_S\bigcup\{(k,n)\}$.  If GN $k$ has overheard file $f_{n}$, i.e., $O_{k,n}=1$, this file placement will not incur additional caching cost, i.e., $\Delta C_U(k,n)=0$. Otherwise, we assume that the UAV needs to visit GN $k$. The minimum distance for UAV to reach GN $k$ from the existing visiting points is approximated as
\begin{align}
{d(k)=\min\{d_{kk'}, k'\in \mathcal{K}_v\}},
\end{align}
where $d_{kk'}$ is the distance between GN $k$ and GN $k'$. Hence, the total additional UAV time required for the UAV to cache file $f_{n}$ at GN $k$ is given by
\begin{align}
{\Delta C_U(k,n)=\frac{d(k)}{V_{\max}}+Yt_p^U},
\end{align}
where $V_{\max}$ is the maximum UAV speed, $Y$ is the number of packets per file and $t_p^U$ is the time required for the UAV to transmit one packet.

Let the GN-file pair $(k^*,n^*)$ be the best candidate in $\mathcal{I}_C$ that leads to the highest net cost reduction for problem \textbf{P2}. If $O_{k^*,n^*}=0$, the UAV needs to visit GN $k^*$ for caching file $f_{n^*}$, and hence we need to update $\mathcal{K}_v$ as $\mathcal{K}_v=\mathcal{K}_v\bigcup\{k^*\}$. Furthermore, we also need to update the overhearing indication variables for the neighbors of GN $k^*$, i.e., $O_{k,n}=1, \forall k\in\mathcal{N}_{k^*}$. On the other hand, if $O_{k^*,n^*}=1$, file $f_{n*}$ is cached at GN $k$ via overhearing, and hence $\mathcal{K}_v$ and $\{O_{k,n}\}$ do not need to be updated.
\begin{example}
Consider a toy example shown in Fig.~\ref{F:example} with $N=3$ and $Q=1$. We assume that the set of selected GN-file pairs is $\mathcal{I}_S=\{(k_1,f_1),(k_2,f_2),(k_3,f_3),(k_4,f_1)\}$. Among the 4 GNs with files, $k_1$, $k_2$ and $k_3$ receive the file by dedicated transmission from UAV, i.e., $\mathcal K_v=\{k_1,k_2,k_3\}$. Next, if we would like to cache $f_3$ at GN $k_6$, it incurs no additional file caching cost since $k_6$ can overhear $f_3$ when the UAV transmits it to $k_3$. On the other hand, if we would like to cache $f_3$ at GN $k_5$, the  additional file caching cost is measured by the time for the UAV to visit $k_5$ from the nearest hovering point, i.e., the location of $k_1$, together with the time for transmitting $f_3$ dedicatedly to $k_5$.
 \begin{figure}[htb]
\centering
\includegraphics[scale=0.7]{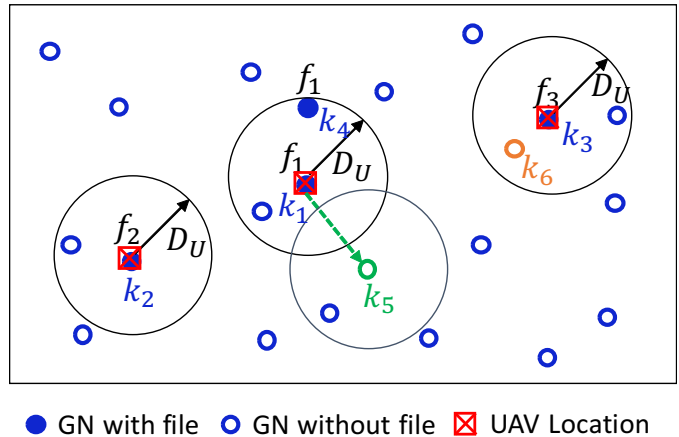}
\caption{A toy example for illustrating the estimation of file caching cost.}
\label{F:example}
\end{figure}
\end{example}

In summary, the pseudo-code for solving problem \textbf{P1} with estimated file caching cost  is summarized in Algorithm~\ref{alg2}.  Since Algorithm~\ref{alg2} does not need solving any optimization problem at the intermediate steps, it can be much more efficiently implemented than Algorithm~\ref{alg1}. Note that Algorithm~\ref{alg1} and Algorithm~\ref{alg2} are identical when $\theta=1$.
\begin{algorithm}
\caption{Greedy Algorithm for Problem \textbf{P1} with estimated file caching cost}
\label{alg2}
\begin{algorithmic}[1]
\STATE{\textbf{Input: }$ \{\mathbf{w}_k,k=1,...,K\},\{P_f(n),n=1,...,N\},Q,D_U.$ }
\STATE{\textbf{Initialize:} Selected GN-file pairs $\mathcal{I}_S=\emptyset$, candidate GN-file pairs $\mathcal{I}_C=\mathcal{I}$, file retrieval cost $C_{G}(\mathcal{I}_S)=\infty$, UAV visiting GNs $\mathcal K_v$, Overhearing variables $\{O_{k,n}=0,\forall k,n\}$ and $\Delta C_{\theta}=\infty$.}
\WHILE{$\Delta C_{\theta}>0$ and $\mathcal{I}_C\neq\emptyset$.}
\FOR{$(k,n)\in\mathcal{I}_C$}
\STATE{Calculate the file retrieval cost $C_G(\mathcal{I}_S\bigcup\{(k,n)\})$ from \eqref{eq:RetrievalCost}.}
\IF{$O_{k,n}=1$}
\STATE{$\Delta C_{U}(k,n)=0$.}
\ELSE
\STATE{$\Delta C_{U}(k,n)=\frac{d(k,n)}{V_{\max}}+Yt_p^U$.}
\ENDIF
\STATE{Net cost reduction: $\Delta C_{\theta}(k,n)=\theta(C_{G}(\mathcal{I}_S)-C_G(\mathcal{I}_S\bigcup\{(k,n)\}))+(1-\theta)\Delta C_{U}(k,n)$.}
\ENDFOR
\STATE{Denote the GN-file pair that leads to the highest cost reduction $\Delta C_{\theta}$ as $(k^*,n^*)$.}
\STATE{Update the selected GN-file pairs $\mathcal{I}=\mathcal{I}\bigcup\{(k^*,n^*)\}$.}
\STATE{Update the candidate GN-file pairs $\mathcal{I}_C$ as in \eqref{eq:updateRemain}.}
\STATE{Update file retrieval cost $C_{G}(\mathcal{I}_S)=C_G(\mathcal{I}_S\bigcup\{(k^*,n^*)\})$.}
\IF{$O_{k^*,n^*}=0$}
\STATE{Update $\mathcal{K}_v=\mathcal{K}_v\bigcup\{k^*\}$ and $O_{k,n}=1, \forall k\in\mathcal{N}_{k^*}$.}
\ENDIF
\ENDWHILE
\STATE{Solve \textbf{P3(a)} with $\mathcal{I}_S$ to obtain the UAV trajectory $\{\mathbf{q}[m]\}$ and file scheduling $\{J_{m,n}\}$.}
\STATE{\textbf{Output}: $\mathcal{I}_S, \{\mathbf{q}[m]\}$ and $\{J_{m,n}\}$}
\end{algorithmic}
\end{algorithm}

\section{Numerical Results}\label{sec:numerical}
In this section, numerical results are provided to evaluate the performance of our proposed scheme for UAV-enabled communication with proactive caching. We assume that the $K$ GNs are randomly and uniformly distributed in a square area of side length equal to $3000$ m and {all the GNs have the same interest on a set of $N$ files}. Unless otherwise stated, the numerical setup of the simulations is given in Table~\ref{tb:numerical}.
\begin{table}[htb]
\centering
\caption{System setup for simulations.}
\begin{tabular}{l|l}
  \hline
  Number of nodes $K$ & 100 \\
  Number of files $N$ & 30 \\
  File size  & 300 kb\\
  GN cache storage $Q$ & 3 \\
  Packet size $R_p$ & 1000 bits \\
   Number of coded packets required  $Y$ & 300  \\
  Zipf distribution parameter $\kappa$ & 1\\
  Channel bandwidth $B_U=B_G$ & 100 kHz \\
  GN transmission power $P_G$ & 20 dBm \\
  GN transmission rate $R_G$ & 10 kb/s \\
  Ground channel fading characteristics & Rayleigh \\
  Ground channel path loss component & 2.7 \\
  SNR gap $\Gamma$ & 7 dB \\
  UAV altitude $H$ & 100 m\\
  UAV to ground path loss component & 2 \\
  UAV transmission power $P_U$ & 10 dBm \\
  UAV transmission rate $R_U$ & 100 kb/s \\
  Maximum UAV speed $V_{\max}$ &  30 m/s \\
  \hline
\end{tabular}
\label{tb:numerical}
\end{table}

 \begin{figure}[htb]
\centering
\includegraphics[scale=0.8]{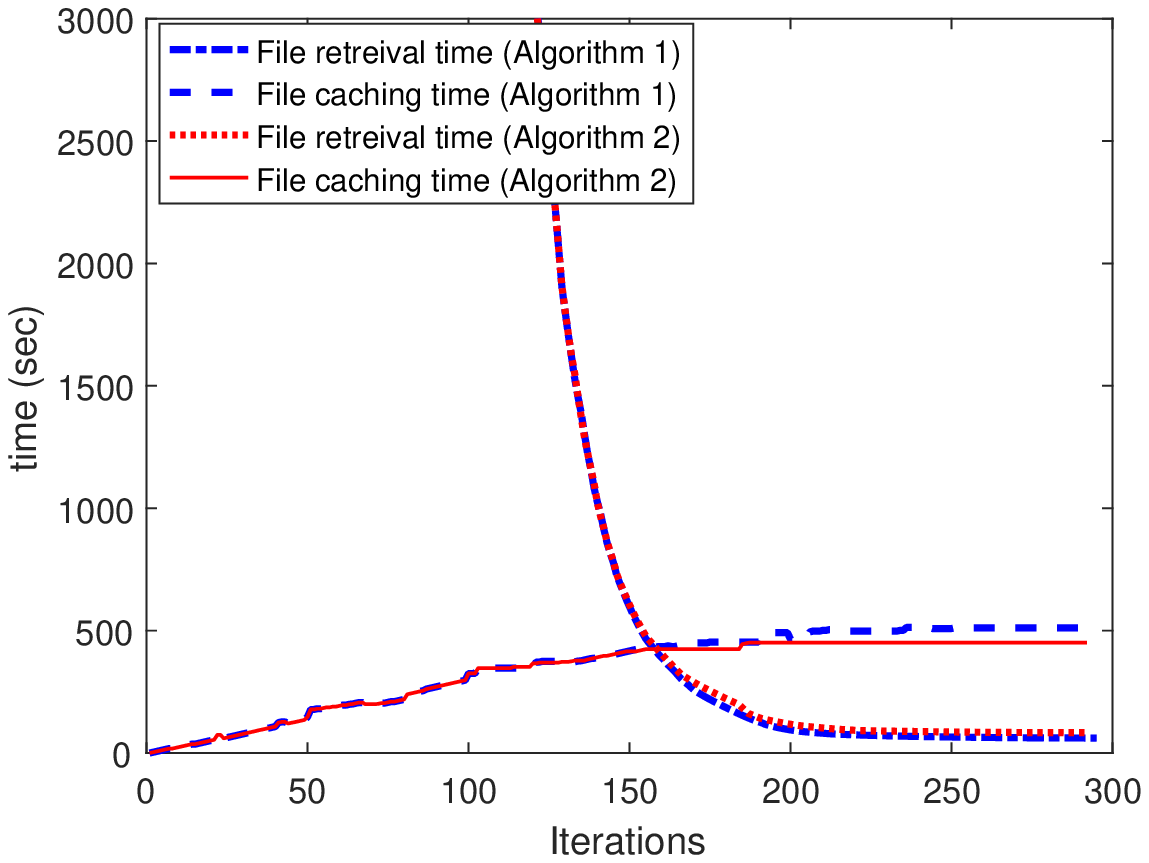}
\caption{The convergence of the file caching and retrieval costs for the proposed greedy algorithms.}
\label{F:convergency}
\end{figure}

Fig.~\ref{F:convergency} shows the convergence behavior of the proposed Algorithm~\ref{alg1} and Algorithm~\ref{alg2} for the example network in Fig.~\ref{F:VBS} with cost weighting factor $\theta=0.6$. It is observed that as the number of selected GN-file pairs (or the number of iterations) increases, the average file retrieval cost decreases, while the UAV caching cost generally increases.  The caching policy returned by Algorithm~\ref{alg1} requires caching time 518.5 s and the resultant average file retrieval time is 54.92 s. On the other hand, the caching policy returned by Algorithm~\ref{alg2} results in file caching time 485.7 s and file retrieval time 73.39 s. Note that Algorithm~\ref{alg2} generally converges to smaller file caching cost but larger file retrieval cost compared with Algorithm~\ref{alg1}. This is as expected, since the estimated file caching cost used for selecting the best GN-file pair in Algorithm~\ref{alg2} is usually larger than the actual file caching cost with optimized UAV trajectory and file transmission scheduling with  Algorithm~\ref{alg1}. Hence, the file caching cost is somewhat over-emphasized by Algorithm~\ref{alg2} in the weighted sum cost, compared to that in Algorithm~\ref{alg1} under  the same cost weighting factor. Furthermore, we compare the resultant file caching policy obtained using Algorithm~\ref{alg1} and Algorithm~\ref{alg2}, denoted by $\{I_{k,n}^{(1)}\}$ and $\{I_{k,n}^{(2)}\}$, respectively. Algorithm~\ref{alg1} and Algorithm~\ref{alg2} terminate after selecting 300 and 292 GN-file pairs, respectively. The discrepancy between these two file caching policies, defined as $\frac{1}{NK}\sum_{n=1}^{N}\sum_{k=1}^{K}|I_{k,n}^{(1)}-I_{k,n}^{(2)}|$, is found to be within 10\%.

 \begin{figure}[htb]
\centering
\includegraphics[scale=0.8]{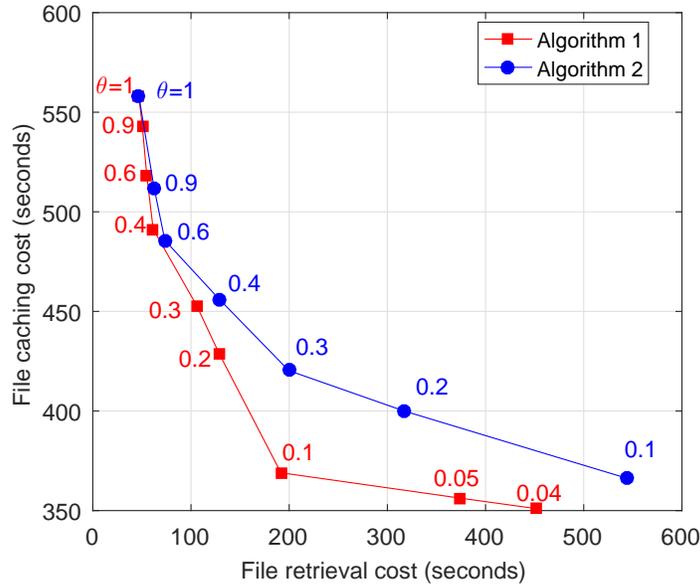}
\caption{File caching and retrieval costs tradeoff.}
\label{F:tradeoff}
\end{figure}

Fig.~\ref{F:tradeoff} depicts the trade-off between the file caching and retrieval costs for the proposed design by using different values for the weighting factor $\theta$. As expected, the file caching cost increases with $\theta$ while the file retrieval cost decreases with $\theta$.  When $\theta=1$, the file caching cost is ignored and the proposed greedy algorithm returns the minimum file retrieval time, as in problem \textbf{P1(a)}, which is $46.9$ s. The corresponding UAV caching time is $558$ s, which is consistent for Algorithm~\ref{alg1} and Algorithm~\ref{alg2}. For $\theta<1$, Algorithm~\ref{alg1} achieves better performance than Algorithm~\ref{alg2}, thanks to the optimization of the UAV trajectory and transmission scheduling at each iteration.  Therefore, in the following, Algorithm~\ref{alg2} is used for performance comparison with other benchmark schemes, since it generally gives a performance lower bound for Algorithm 1 with reduced complexity.

 \begin{figure}[htb]
\centering
\includegraphics[scale=0.8]{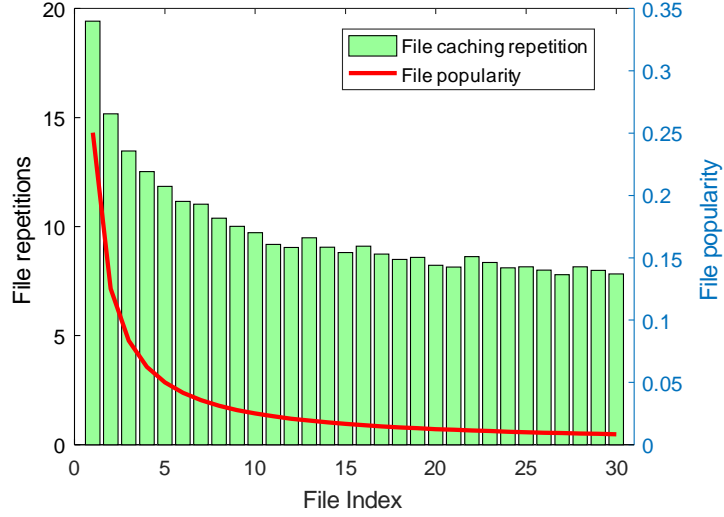}
\caption{File caching repetition versus file popularity.}
\label{F:FileRepeat}
\end{figure}

In Fig.~\ref{F:FileRepeat}, the number of caching repetitions for each file, namely the number of caching GNs that have cached the same file, is plotted together with the file popularity. The simulation results are obtained by averaging over 100 random realizations of the GN locations. As expected, the more popular files are in general cached by more GNs so that they can be retrieved with higher chance from local cache or shorter distances for D2D transmission. Note that the locations of the caching GNs also affect the average file retrieval distance. Hence, we observe that the number of caching repetitions does not necessarily strictly decrease with the file popularity.

Last, to further illustrate the performance gain of the proposed joint design of caching policy, UAV trajectory and transmission scheduling, the following two benchmark schemes are considered:
\begin{itemize}
\item{{\it Benchmark scheme 1}: In this scheme, only the UAV trajectory is optimized, while the caching policy is determined based on the {\it random proportional caching strategy} \cite{Ji2015}. Specifically, each GN selects its caching file by independently generating the random file indices with probability proportional to the file popularity $P_f(n)$. Then, with the file caching policy obtained, the proposed UAV trajectory and transmission scheduling optimization in Section~\ref{sec:general} are applied for the UAV to transmit all the files to their designated caching GNs. }
\item{{\it Benchmark scheme 2}: In this scheme, only the file caching policy is optimized and the UAV trajectory follows the conventional TSP solution with the selected caching GNs. Specifically, the caching policy is determined by minimizing the file retrieval cost via solving \textbf{P1(a)}. Then, the UAV flying path is designed by solving the conventional TSP with the caching GNs as waypoints. The transmission scheduling design follows the proposed scheme in Section~\ref{sec:scheduling}.}
\item{{\it Benchmark scheme 3}: In this scheme, the file placement follows the optimal probabilistic caching policy proposed in \cite{Blaszczyszyn2015}, and the UAV trajectory is optimized based on the proposed algorithm. Specifically, each GN selects its caching file by independently generating the random file indices with probabilities that are optimized to minimize the file retrieval cost. Then, with the file caching policy obtained, the proposed UAV trajectory and transmission scheduling optimization are applied are applied for the UAV to transmit all the files to their designated caching GNs.}
\end{itemize}

\begin{figure}[htb]
\centering
\includegraphics[scale=0.9]{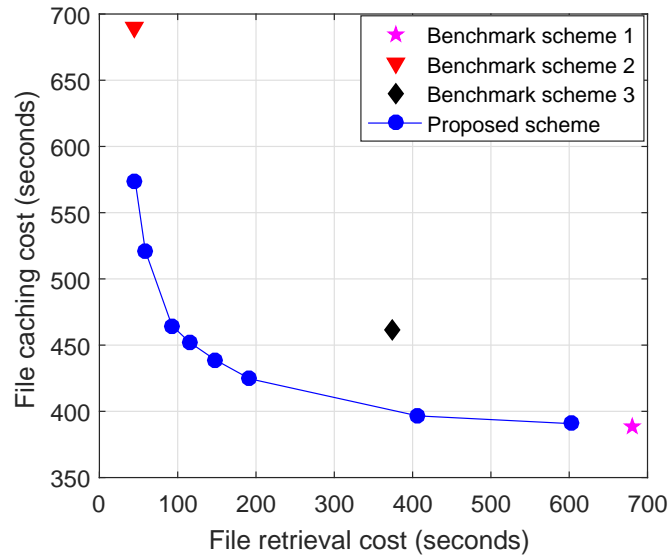}
\caption{Performance comparison of the proposed scheme with benchmark schemes.}
\label{F:compare}
\end{figure}

The simulation results obtained by averaging over 100 random realizations of the GN locations are shown in Fig.~\ref{F:compare}. It is firstly observed that each of the two benchmark schemes only result in one singleton cost trade-off point in Fig.~\ref{F:compare}. This is expected since they either optimize the file caching cost or the file retrieval cost only, thus no flexible trade-off between the two costs can be achieved, as in our proposed joint design. It is also observed that benchmark scheme 1 with \emph{random proportional caching strategy} leads to comparable minimal  file caching cost as the proposed joint design scheme with $\theta=0.07$, but it results in higher file retrieval cost. { Benchmark scheme 3 with the optimal probabilistic caching policy achieves lower file retrieval cost than the random proportional caching strategy. However, it still under-performs the proposed caching strategy given the same amount of file caching cost. }This illustrates the performance gain of the proposed caching policy over the random caching strategy. On the other hand, the second benchmark scheme with optimized file caching policy by solving \textbf{P1(a)} achieves the same minimum file retrieval cost as the proposed scheme, but it requires much higher file caching cost as compared with the proposed scheme. This is mainly attributed to the performance gain brought by the optimized UAV trajectory design.

\section{Conclusion}\label{sec:con}
In this paper, we have proposed a novel scheme to overcome the issue of limited UAV endurance for UAV-enabled wireless communications, by utilizing the promising technique of proactive caching.  With the proposed scheme, the UAV is only required in the file caching phase while it can replenish its energy during the file retrieval phase, thus resolving the endurance issue.  We have investigated the fundamental trade-off between the file caching cost and retrieval cost for the proposed scheme, by formulating and solving the optimization problem to minimize the weighted sum of the two costs, via jointly designing the file caching policy, UAV trajectory and UAV file transmission scheduling. In practical applications, the weighting factor can be flexibly set to achieve a good balance between affordable UAV endurance and D2D transmission delay/channel occupation according to the system requirement. The extension of the proposed scheme to multiple UAVs and/or multiple file sets of interest  will be left for future work.

\bibliographystyle{ieeetr}
\bibliography{UAV}

\end{document}